\documentclass[preprint,aps,prl,endfloats]{revtex4-1}

\usepackage{graphicx}
\usepackage{dcolumn}
\usepackage{bm}
\usepackage{amsmath}
\usepackage{latexsym}
\usepackage{hyperref}

\newcommand{\rfig}[1]{Fig.~\ref{#1}}
\newcommand{\rref}[1]{Ref.~\onlinecite{#1}}
\newcommand{\req}[1]{Eq.~(\ref{#1})}

\makeindex

\begin{document}

\title{Electrical control of intervalley scattering in graphene via the charge state of defects}

\author{Baoming Yan}
\author{Qi Han}
\author{Zhenzhao Jia}
\author{Jingjing Niu}
\author{Tuocheng Cai}
\author{Dapeng Yu}
\author{Xiaosong Wu}
\email{xswu@pku.edu.cn}
\affiliation{State Key Laboratory for Artificial Microstructure and Mesoscopic Physics\\ Peking University, Beijing, 100871, China \\ Collaborative Innovation Center of Quantum Matter, Beijing 100871, China}

\begin{abstract}
We study the intervalley scattering in defected graphene by low-temperature transport measurements. The scattering rate is strongly suppressed when defects are charged. This finding highlights ``screening'' of the short-range part of a potential by the long-range part. Experiments on calcium-adsorbed graphene confirm the role of a long-range Coulomb potential. This effect is applicable to other multivalley systems, provided that the charge state of a defect can be electrically tuned. Our result provides a means to electrically control valley relaxation and has important implications in valley dynamics in valleytronic materials.
\end{abstract}



\maketitle
\graphicspath{{figures/}}

The emerging so called valleytronics has attracted enormous interest \cite{Rycerz2007,Xu2014}. In analogy to spintronics, where the spin degree of freedom is utilized to encode information, valleytronics exploits the valley degree of freedom. Obviously, valley polarization is the prerequisite for functioning of valleytronic devices. Early studies in graphene have proposed methods for producing polarization, such as edge engineering \cite{Rycerz2007} and breaking the inversion symmetry \cite{Xiao2007}, whereas there was little experimental progress. Transition metal dichalcogenides (TMDs) have particularly intensified the study of valleytronics, because their valley polarization can be readily achieved by optical pumping \cite{Mak2012,Zeng2012}. Very recently, valley polarization in graphene has also been experimentally demonstrated \cite{Gorbachev2014,Ju2015,Sui2015}. In light of all this progress, it becomes increasingly desirable to understand valley dynamics. Valley relaxation takes place via intervalley scattering. Because of the large separation between valleys in the momentum space, short-range potentials are much more effective in generating intervalley scattering. Reducing intervalley scattering by short-range potentials is one of the main approaches in enhancement of the valley coherence time \cite{Wu2013}.

In practical materials, long-range Coulomb potentials are almost always present, in addition to short-range potentials. Previous studies in graphene have considered the case in which two types of potential originate from different sources and independently contribute to scattering \cite{Nomura2007,Adam2007,Hwang2007,Chen2009,Wehling2010,Peres2010}. However, a potential sometimes consists of both a long-range part and a short-range part. Study on the interplay of these two has been missing so far. In this Rapid Communication, we study the effect of the long-range potential on short-range scattering, e.g., intervalley scattering. We choose graphene as an example material for two reasons. First, it is well established in graphene that the intervalley scattering time can be extracted from weak-localization magnetoresistance \cite{Mccann2006,Wu2007,Tikhonenko2008}. Second, because defects lead to resonant states at the Dirac point \cite{Pereira2006,Wehling2007}, the charge state, hence the Coulomb potential, can be turned on and off simply by tuning the gate voltage \cite{Brar2011}. This provides a substantial advantage in studying the influence of the Coulomb potential. The main result is that when defects are charged, intervalley scattering is strongly suppressed. The effect is rather general and expected in other materials as well. It is particularly important in two-dimensional materials in which Coulomb interaction is strongly enhanced. Our study offers a perspective on intervalley scattering in valleytronic materials.

Graphene flakes were exfoliated from Kish graphite onto 285-nm SiO$_2$/Si substrates. To obtain an ultra-clean surface, a shadow mask method was used to fabricate graphene devices. Electrodes were made of 5 nm Pd/ 80 nm Au by e-beam deposition. Point defects were created by bombardment with low-density Ar$^+$ in a standard reactive-ion etching system \cite{Chen2013}. The density of point defects was estimated by Raman spectroscopy. Calcium deposition onto graphene was carried out by \emph{in situ} thermal deposition in an ultralow-temperature transport measurement apparatus. In order to suppress adatom diffusion and clustering on graphene, the sample temperature throughout deposition and measurements was maintained below 10 K, which is much less than the diffusion barrier for calcium on graphene \cite{Nakada2011}. The samples are expected in ultra high vacuum owing to the strong cryopumping effect. After the samples were cooled down to helium temperature, current annealing (current density was 0.2 mA/$\mu$m) was performed to remove possible gas adsorption. Electrical measurements were performed by a standard low-frequency lock-in technique. Raman measurements were performed, immediately after transport measurements, on a Renishaw InVia micro-Raman system with 514 nm laser excitation.

\begin{figure}[htbp]
	\begin{center}
		\includegraphics[width=0.9\columnwidth]{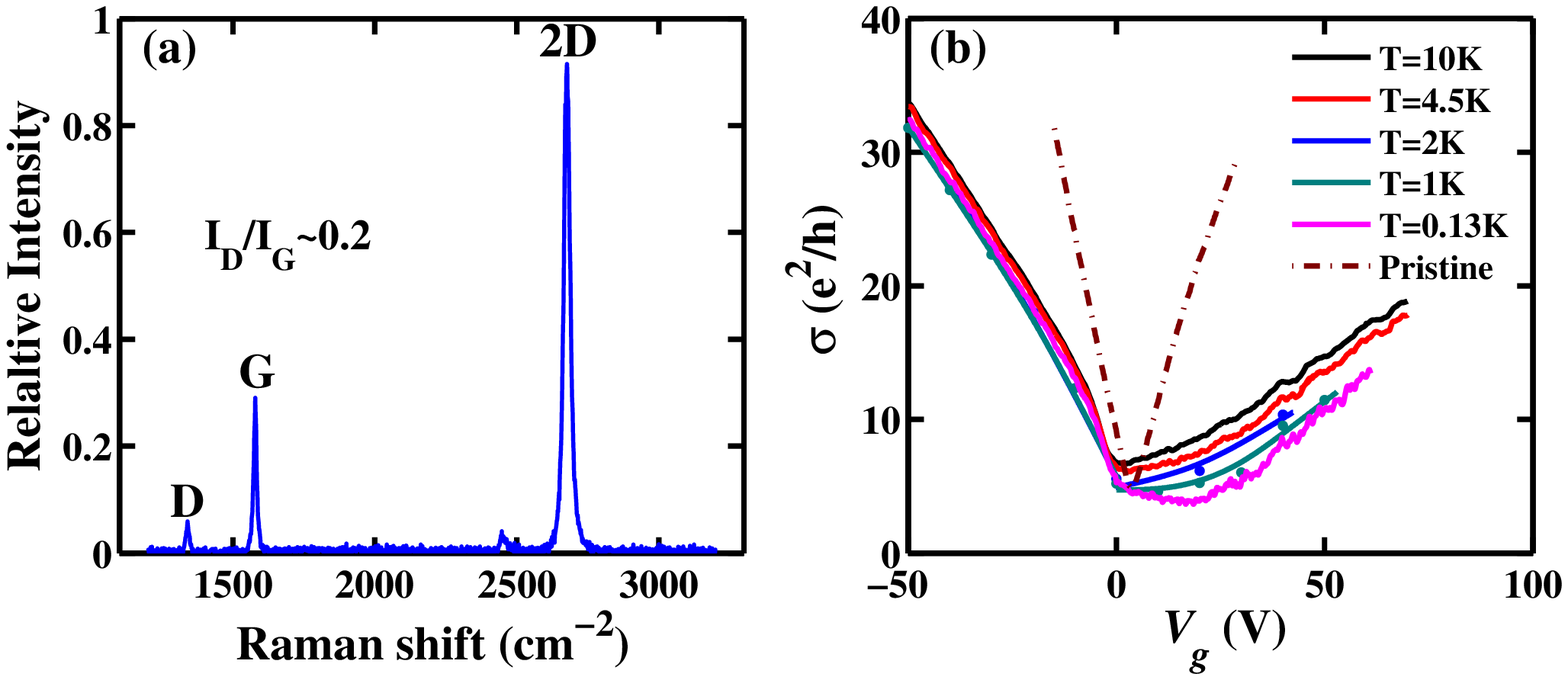}
		\caption{(a) Raman spectrum measured after introducing defects by Ar$^+$ bombardment. Pronounced $D$ peak is observed with $I_D/I_G\sim 0.2$, corresponding to a defect concentration of $1.9\times10^{11}$ cm$^{-2}$. (b) Conductivity vs gate voltage at different temperatures. For comparison, the conductivity of a typical pristine graphene sample at 4.5 K is plotted as a dashed line. Separated data points at 1 K and 2 K are taken from the magnetoconductivity data at $B=0$ and linked by lines as a guide to the eye.}
		\label{s1rvgt}
	\end{center}
\end{figure}

The Raman spectrum of a defected graphene sample (by Ar ion bombardment) is shown in \rfig{s1rvgt}(a). Besides the characteristic $G$ and $2D$ peaks, the spectrum consists of a $D$ peak at about 1350 cm$^{-1}$, which is a signature of point defects. The defect density can be estimated according to an empirical formula developed by Lucchese \emph{et al}. \cite{Lucchese2010}:
 \begin{equation}\label{eq_raman}
\frac{C(\lambda)}{L_D^2}=\frac{I_D}{I_G}, \mbox{\,with\,\,} C(\lambda=514 \textrm{nm})\sim 107 \textrm{nm}^2
\end{equation}
where $I_D$ ($I_G$) is the intensity of the $D$ ($G$) peak, and $L_D$ is the average distance between two defects. $L_D$ turns out to be $\sim23$ nm. It corresponds to a defect concentration of $1.9\times10^{11}$ cm$^{-2}$.

We now turn to the gate voltage dependence of conductivity of defected graphene. For typical pristine graphene, the dependence displays a slight asymmetry with respect to the Dirac point, which is usually ascribed to the difference between scattering off an attractive potential and off a repulsive one \cite{Novikov2007}. However, in defected graphene, a strong asymmetry appears [see \rfig{s1rvgt}(b)]. The slope on the hole side (negative gate voltage) is substantially larger than that on the electron side (positive gate voltage). The field effect mobilities for hole and electron are 1650 cm$^2$/V$\cdot$s ($\mu_h$) and 738 cm$^2$/V$\cdot$s ($\mu_e$), respectively. The resultant ratio is $\mu_e/\mu_h\sim0.45$, appreciably smaller than that of typical pristine graphene and potassium doped graphene \cite{Chen2008}. Furthermore, the evolution of the conductance is different for holes and electrons. With increase of the hole density, a steep increase of the conductivity with the gate voltage in the vicinity of the Dirac point is followed by a weaker linear dependence at high carrier density. On the contrary, for electrons, the conductivity undergoes a slow increase and then develops a larger slope. The most striking feature is that the temperature dependence of the conductivity displays distinctive behavior for holes and electrons, i.e., virtually independent of $T$ for holes, while strongly $T$ dependent for electrons. These unusual behaviors suggest that scattering mechanism is fundamentally different in the two regions.

\begin{figure}[htp]
	\begin{center}
		\includegraphics[width=1\columnwidth]{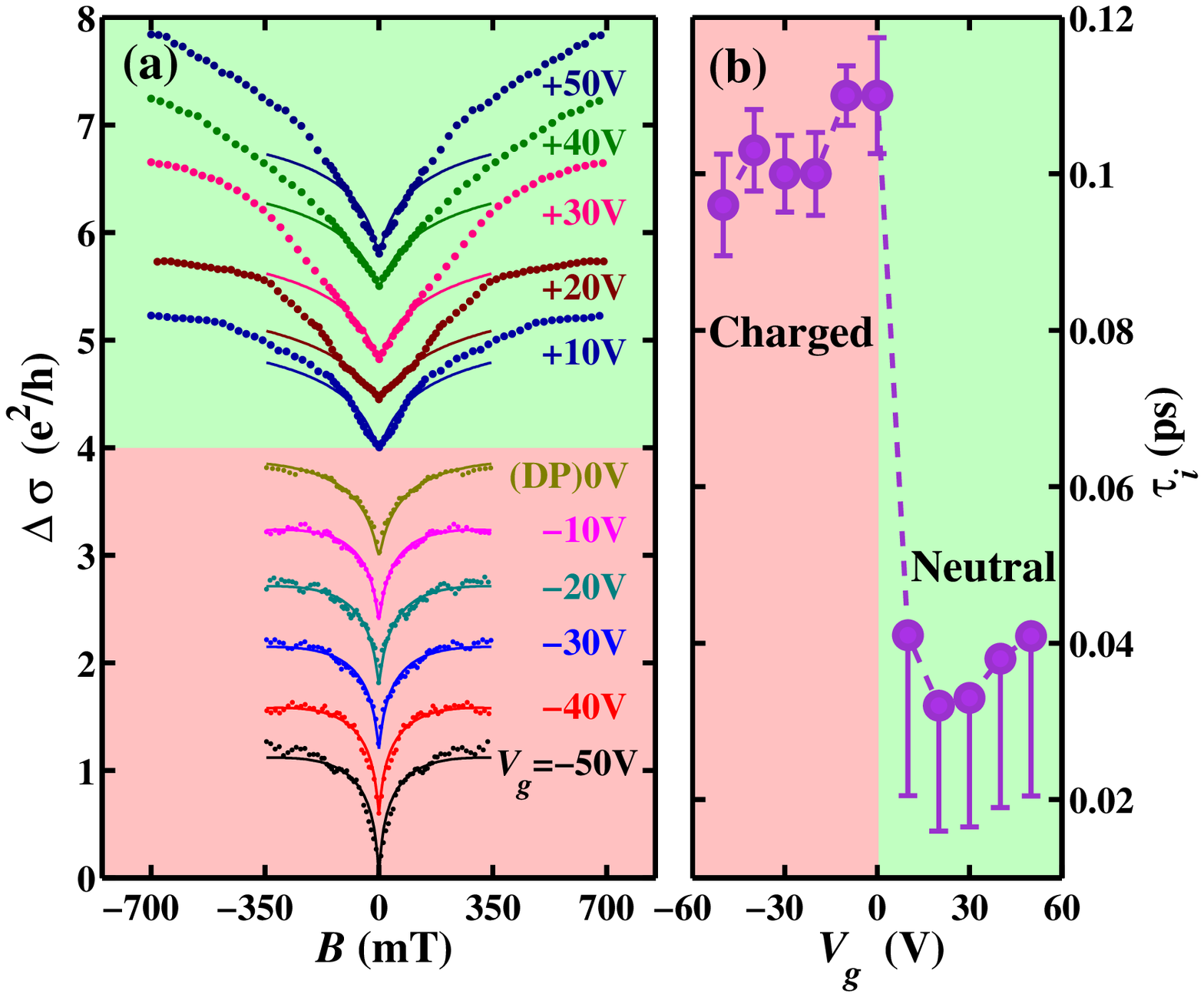}
		\caption{Intervalley scattering in defected graphene. (a) Low-field magnetoconductivity $\Delta \sigma$ of defected graphene (dots) at various gate voltages at $T=1$ K. Data are shifted in $y$ for clarity. Solid lines are best fits of the experimental data to \req{eq_gwl}. Different colors of shading represent the different charge states of defects (see the text), light pink for the charged state and light green for the neutral state. (b) Intervalley scattering times obtained by fitting are plotted as a function of the gate voltage. On the charged side, the error bars represent the uncertainty in the fitting to \req{eq_gwl}, while on the neutral side, they represent the upper bound (2$\tau$; see the text) and the lower bound ($\tau$) of $\tau_i$.}
		\label{s1rbvg}
	\end{center}
\end{figure}

Magnetoconductivity data support different scattering mechanisms in the two regions. The change of the conductivity with the magnetic field, $\Delta\sigma(B)=\sigma(B)-\sigma(0)$, is depicted in \rfig{s1rbvg}(a). In the hole region, a narrow dip appears around $B=0$. However, in the electron region, it becomes much broader. The dip structure originates from weak localization. It is a direct consequence of intervalley scattering \cite{Mccann2006}. It is well known that the interference of electron waves along time-reversal paths gives rise to a quantum correction to the conductivity. A magnetic field suppresses the correction by randomizing the phases of interfering waves. In intrinsic graphene, the inference is destructive because of a Berry phase of $\pi$. As a result, the magnetoconductivity is negative, i.e., weak antilocalization(WAL). In the presence of intervalley scattering, the magnetoconductivity in low fields turns positive, i.e.,weak localization (WL). The width of the WL dip is directly related to the strength of intervalley scattering. Roughly, the width $\Delta B$ is inversely proportional to the intervalley scattering time $\tau_i$, $\Delta B \approx \hbar/2eD\tau_i$. Here, $e$ is the elementary charge, $\hbar$ the reduced Planck constant, and $D$ the diffusion constant. So, the data in \rfig{s1rbvg}(a) clearly show that intervalley scattering rate $1/\tau_i$ is much larger on the electron side than on the hole side.

Quantitative estimation of $\tau_i$ can be obtained by fitting the experimental data to an equation widely used to describe WAL in graphene \cite{Mccann2006}:
\begin{equation}\label{eq_gwl}
\begin{split}
\delta\sigma(B)=&\frac{e^2}{\pi h}\Bigg[F\left(\frac{B_z}{B_\phi}\right)-F\left(\frac{B_z}{B_\phi+2B_i}\right)\\ &\,\,\,\,\,\,\,-2F\left(\frac{B_z}{B_\phi+B_i+B_*}\right)\Bigg]\\
F(z)&=\ln(z)+\psi\left(\frac{1}{2}+\frac{1}{z}\right),B_{\phi,i,*}=\frac{\hbar}{4De\tau_{\phi,i,*}},
\end{split}
\end{equation}
where $\psi$ is the digamma function, $\tau_\phi$ the phase coherence time, and $\tau_*$ the intravalley scattering time. The fitted parameter $\tau_i$ as a function of $V_g$ is plotted in \rfig{s1rbvg}(b). On the hole side, $\tau_i$ is about 0.1 ps. After the Fermi level passes the Dirac point towards the electron side, it suddenly drops to 0.04 ps, which is the lower bound set for fitting. In the theory of WL, electrons are assumed to be diffusive, \emph{i.e.} $\tau_{\phi,i,*}\gg \tau$. Here, $\tau$ is the mean-free time, which can be calculated from the conductivity. To satisfy the assumption, $2\tau$ was set as the lower bound of the fitting parameter $\tau_i$. The fact that $\tau_i$ reached the bound indicates dominance of intervalley scattering by defects, which is consistent with the average distance of 23 nm between defects inferred from the Raman spectrum. The deviation of data from fitting curves is most likely due to the assumption of \req{eq_gwl} being no longer held, which implies that the actual $\tau_i$ is less than $2\tau$. Nevertheless, it is unambiguous that the intervalley scattering rate $1/\tau_i$ is strongly suppressed on the hole side.

Intervalley scattering is caused by atomically sharp short-range potentials, such as defects. However, the defect concentration will apparently not change with $V_g$. The only plausible explanation is that a mechanism prevents carriers from ``seeing'' defects. The electronic structure of defects in graphene has been theoretically studied \cite{Pereira2006,Wehling2007,Nanda2012}. There is a general consensus that defects introduce resonant states at the Dirac point \cite{Note1}. The resonance state has been confirmed experimentally \cite{Ugeda2010}. When the Fermi level passes the Dirac point, the resonant states will be populated/unpopulated, which corresponds to neutral/charged states of defects. This gate-controlled charging effect for defects has been demonstrated in a scanning tunneling microscopy study \cite{Brar2011}, in which not only a defect resonance state has been observed, but ionization of a defect by tip induced gating has been realized. A recent experimental study, combined with theoretical calculations, has also shown that defects are positively charged \cite{Liu2014a}. Therefore, on the hole side, where defects are ionized (charged), the long-range Coulomb potential can deflect carriers and reduce the scattering cross section of the short-range potential, leading to suppression of intervalley scattering.

\begin{figure}[htbp]
	\begin{center}
		\includegraphics[width=0.9\columnwidth]{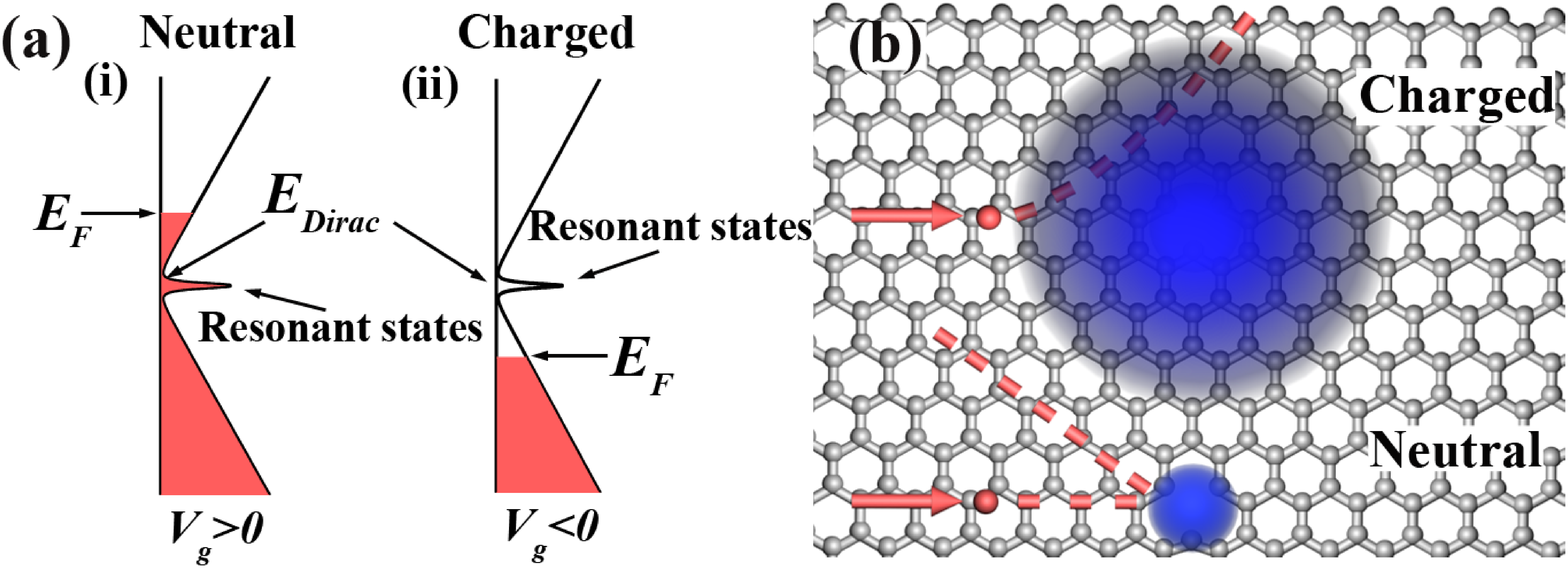}
		\caption{Schematic diagram for defect scattering. (a) Density of states of defected graphene. (i) When the Fermi level is in the conduction band (positive gate voltage), defects are neutral, (ii) while they become charged when the Fermi level is in the valance band (negative gate voltage). (b) Carriers scattered off a neutral defect and a charged one.}
		\label{s1_defects}
	\end{center}
\end{figure}

The proposed mechanism is illustrated in \rfig{s1_defects}. Impurity resonant states form at the Dirac point. When the Fermi level is above the Dirac point, impurity states are filled and defects are neutral. Carriers can approach defects and be scattered by short-range potential. Significant intervalley scattering leads to strong weak localization, which is temperature dependent. In contrast, when the Fermi level is below the Dirac point, impurity states are empty and defects are charged. Long-range Coulomb potentials dominate carrier scattering. Few carriers can penetrate the Coulomb potential and be scattered by the short-range potential, resulting in suppression of intervalley scattering. Since the short-range potential contributes more to momentum relaxation than the long-range one in graphene, the mobility exhibits substantial asymmetry.

Around the Dirac point, when the gate voltage is swept from positive to negative, defects go through a transition from a neutral to a charged state. The scattering cross section of the short-range potential continuously decreases because it is screened by the long-range Coulomb potential of charged defects. The total scattering cross section is also reduced as the short-range potential is more effective in scattering carriers than the long-range one, indicated by the asymmetry of the conductivity. On the other hand, the carrier density decreases first and then increases after passing the Dirac point. The combination of the two effects leads to a weak gate dependence of conductivity on the right side of the Dirac point, while to a strong increase of conductivity on the left side, which is in excellent agreement with the experimentally observed evolution of the conductivity near the Dirac point.

\begin{figure}[hbp]
	\begin{center}
		\includegraphics[width=0.9\columnwidth]{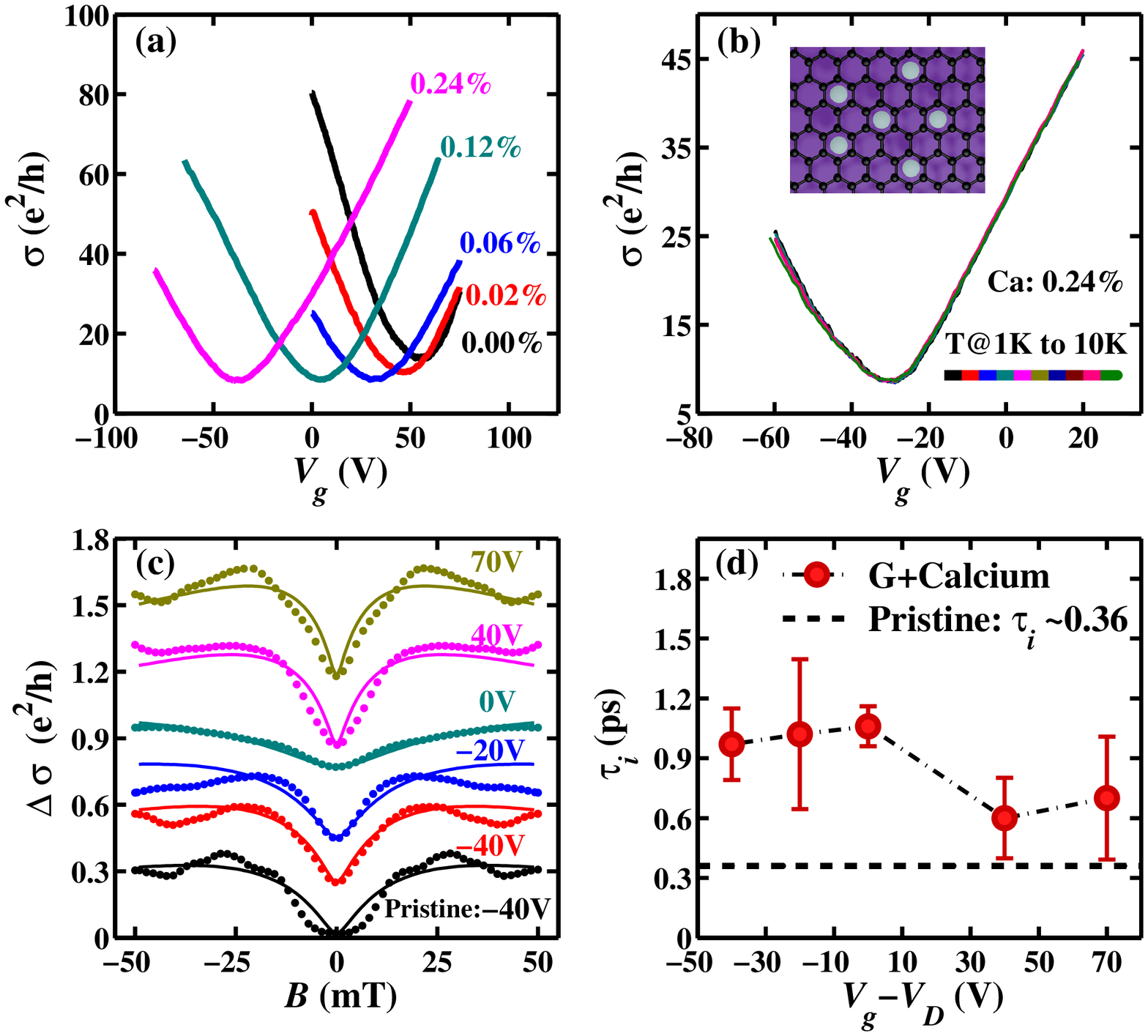}
		\caption{Transport experiments of Ca-doped graphene. (a) Conductivity versus gate voltage at increasing coverage of Ca adatoms. Different colors denote different Ca coverage. (b) The gate dependence of conductivity for a Ca coverage of 0.24\% at temperatures from 1 K to 10 K at steps of 1 K. Inset: Schematic diagram of Ca adatoms on graphene. Ca adatoms are assumed to be at the hollow site\cite{Chan2008,Nakada2011}. (c) Low-field magnetoconductivity data (dots) for a Ca coverage of 0.24\% at various gate voltages from -40 to 70 V relative to the Dirac point voltage ($T$=150 mK). The black curve is the conductivity before Ca deposition (pristine). Solid lines are best fits of experimental data to \req{eq_gwl}. (d) Intervalley scattering time $\tau_i$ obtained by fitting to \req{eq_gwl} is plotted as a function of gate voltage. The black dashed line denotes $\tau_i$ of pristine graphene in (c).}
		\label{s2tauvg}
	\end{center}
\end{figure}

To further confirm the suppression of intervalley scattering by Coulomb potential, transport experiments were performed on calcium-doped graphene. According to previous theoretical calculations, Ca adatoms form ionic bonds to graphene and are only $\sim2.3$ {\AA} away from the graphene sheet \cite{Chan2008}. The charge transfer per Ca adatom is predicted to be 0.78$e$\cite{Chan2008}. So, the potential should have both a short-range part and a long-range part, just like a charged defect. This should hold even if the actual bond length may deviate from the 2.3 {\AA}, as it only yields variation of the strength of the short-range part. The transport data are shown in \rfig{s2tauvg}. Upon Ca deposition, the Dirac point shifts towards the negative gate voltage, suggesting electron doping. Note that Ca adatoms stay in a positively charged state in the whole range of gate voltage. No apparent temperature dependence was observed, consistent with defected graphene when defects are charged. Mobility decreases due to additional scattering by adatoms. Interestingly, the weak-localization peak in the magnetoconductivity became slightly \emph{narrower} after Ca deposition. Note that the peak should be \emph{wider} when the mobility is reduced, even if the intervalley scattering is not changed. So, not only did Ca adatoms not introduce intervalley scattering, but the intervalley scattering was suppressed. By fitting data to \req{eq_gwl}, we obtained the intervalley scattering time, which indeed increased from 0.36 to about 0.8 ps. Thus, absence of enhancement of intervalley scattering in Ca-doped graphene corroborates with suppressed intervalley scattering in defected graphene when defects are charged.

In conclusion, our experiment unambiguously shows that the long-range part of a potential can strongly reduce the transport cross section of the short-range part. In a multivalley system, this means suppression of intervalley scattering. Therefore, this effect is particularly relevant in materials promising for valleytronics, such as graphene and TMDs. Furthermore, we would like to point out that it also provides important insight into another topic in which adatoms have been proposed to engineer the properties of graphene, such as spin-orbit coupling \cite{CastroNeto2009,Weeks2011,Jiang2012,Ferreira2014,Cresti2014} and magnetism \cite{Sevincli2008,Qiao2010,Cao2010,Virgus2014}. Despite many theoretical proposals, transport studies have shown that Au, Mg and In adatoms on graphene generate only long-range Coulomb impurity scattering, while neither spin-orbit scattering nor magnetic scattering has been introduced \cite{Pi2010,Swartz2013,Jia2015,Chandni2015}. Since all these adatoms are charged, our result points to a reasonable explanation and a hint for possible solutions.

\begin{acknowledgements}
This work was supported by the National Key Basic Research Program of China (No. 2012CB933404 and No. 2013CBA01603) and NSFC (Projects No. 11074007, No. 11222436, and No. 11234001).
\end{acknowledgements}


\begin{thebibliography}{44}
\expandafter\ifx\csname natexlab\endcsname\relax\def\natexlab#1{#1}\fi
\expandafter\ifx\csname bibnamefont\endcsname\relax
  \def\bibnamefont#1{#1}\fi
\expandafter\ifx\csname bibfnamefont\endcsname\relax
  \def\bibfnamefont#1{#1}\fi
\expandafter\ifx\csname citenamefont\endcsname\relax
  \def\citenamefont#1{#1}\fi
\expandafter\ifx\csname url\endcsname\relax
  \def\url#1{\texttt{#1}}\fi
\expandafter\ifx\csname urlprefix\endcsname\relax\def\urlprefix{URL }\fi
\providecommand{\bibinfo}[2]{#2}
\providecommand{\eprint}[2][]{\url{#2}}

\bibitem[{\citenamefont{Rycerz et~al.}(2007)\citenamefont{Rycerz, Tworzydlo,
  and Beenakker}}]{Rycerz2007}
\bibinfo{author}{\bibfnamefont{A.}~\bibnamefont{Rycerz}},
  \bibinfo{author}{\bibfnamefont{J.}~\bibnamefont{Tworzydlo}},
  \bibnamefont{and} \bibinfo{author}{\bibfnamefont{C.~W.~J.}
  \bibnamefont{Beenakker}}, \bibinfo{journal}{Nature Phys.}
  \textbf{\bibinfo{volume}{3}}, \bibinfo{pages}{172} (\bibinfo{year}{2007}).

\bibitem[{\citenamefont{Xu et~al.}(2014)\citenamefont{Xu, Yao, Xiao, and
  Heinz}}]{Xu2014}
\bibinfo{author}{\bibfnamefont{X.}~\bibnamefont{Xu}},
  \bibinfo{author}{\bibfnamefont{W.}~\bibnamefont{Yao}},
  \bibinfo{author}{\bibfnamefont{D.}~\bibnamefont{Xiao}}, \bibnamefont{and}
  \bibinfo{author}{\bibfnamefont{T.~F.} \bibnamefont{Heinz}},
  \bibinfo{journal}{Nat Phys} \textbf{\bibinfo{volume}{10}},
  \bibinfo{pages}{343} (\bibinfo{year}{2014}).

\bibitem[{\citenamefont{Xiao et~al.}(2007)\citenamefont{Xiao, Yao, and
  Niu}}]{Xiao2007}
\bibinfo{author}{\bibfnamefont{D.}~\bibnamefont{Xiao}},
  \bibinfo{author}{\bibfnamefont{W.}~\bibnamefont{Yao}}, \bibnamefont{and}
  \bibinfo{author}{\bibfnamefont{Q.}~\bibnamefont{Niu}},
  \bibinfo{journal}{Phys. Rev. Lett.} \textbf{\bibinfo{volume}{99}},
  \bibinfo{eid}{236809} (\bibinfo{year}{2007}).

\bibitem[{\citenamefont{Mak et~al.}(2012)\citenamefont{Mak, He, Shan, and
  Heinz}}]{Mak2012}
\bibinfo{author}{\bibfnamefont{K.~F.} \bibnamefont{Mak}},
  \bibinfo{author}{\bibfnamefont{K.}~\bibnamefont{He}},
  \bibinfo{author}{\bibfnamefont{J.}~\bibnamefont{Shan}}, \bibnamefont{and}
  \bibinfo{author}{\bibfnamefont{T.~F.} \bibnamefont{Heinz}},
  \bibinfo{journal}{Nat. Nanotechnol.} \textbf{\bibinfo{volume}{7}},
  \bibinfo{pages}{494} (\bibinfo{year}{2012}).

\bibitem[{\citenamefont{Zeng et~al.}(2012)\citenamefont{Zeng, Dai, Yao, Xiao,
  and Cui}}]{Zeng2012}
\bibinfo{author}{\bibfnamefont{H.}~\bibnamefont{Zeng}},
  \bibinfo{author}{\bibfnamefont{J.}~\bibnamefont{Dai}},
  \bibinfo{author}{\bibfnamefont{W.}~\bibnamefont{Yao}},
  \bibinfo{author}{\bibfnamefont{D.}~\bibnamefont{Xiao}}, \bibnamefont{and}
  \bibinfo{author}{\bibfnamefont{X.}~\bibnamefont{Cui}}, \bibinfo{journal}{Nat.
  Nanotechnol.} \textbf{\bibinfo{volume}{7}}, \bibinfo{pages}{490}
  (\bibinfo{year}{2012}).

\bibitem[{\citenamefont{Gorbachev et~al.}(2014)\citenamefont{Gorbachev, Song,
  Yu, Kretinin, Withers, Cao, Mishchenko, Grigorieva, Novoselov, Levitov
  et~al.}}]{Gorbachev2014}
\bibinfo{author}{\bibfnamefont{R.~V.} \bibnamefont{Gorbachev}},
  \bibinfo{author}{\bibfnamefont{J.~C.~W.} \bibnamefont{Song}},
  \bibinfo{author}{\bibfnamefont{G.~L.} \bibnamefont{Yu}},
  \bibinfo{author}{\bibfnamefont{A.~V.} \bibnamefont{Kretinin}},
  \bibinfo{author}{\bibfnamefont{F.}~\bibnamefont{Withers}},
  \bibinfo{author}{\bibfnamefont{Y.}~\bibnamefont{Cao}},
  \bibinfo{author}{\bibfnamefont{A.}~\bibnamefont{Mishchenko}},
  \bibinfo{author}{\bibfnamefont{I.~V.} \bibnamefont{Grigorieva}},
  \bibinfo{author}{\bibfnamefont{K.~S.} \bibnamefont{Novoselov}},
  \bibinfo{author}{\bibfnamefont{L.~S.} \bibnamefont{Levitov}},
  \bibnamefont{et~al.}, \bibinfo{journal}{Science}
  \textbf{\bibinfo{volume}{346}}, \bibinfo{pages}{448} (\bibinfo{year}{2014}).

\bibitem[{\citenamefont{Ju et~al.}(2015)\citenamefont{Ju, Shi, Nair, Lv, Jin,
  Velasco, Ojeda-Aristizabal, Bechtel, Martin, Zettl et~al.}}]{Ju2015}
\bibinfo{author}{\bibfnamefont{L.}~\bibnamefont{Ju}},
  \bibinfo{author}{\bibfnamefont{Z.}~\bibnamefont{Shi}},
  \bibinfo{author}{\bibfnamefont{N.}~\bibnamefont{Nair}},
  \bibinfo{author}{\bibfnamefont{Y.}~\bibnamefont{Lv}},
  \bibinfo{author}{\bibfnamefont{C.}~\bibnamefont{Jin}},
  \bibinfo{author}{\bibfnamefont{J.}~\bibnamefont{Velasco}},
  \bibinfo{author}{\bibfnamefont{C.}~\bibnamefont{Ojeda-Aristizabal}},
  \bibinfo{author}{\bibfnamefont{H.~A.} \bibnamefont{Bechtel}},
  \bibinfo{author}{\bibfnamefont{M.~C.} \bibnamefont{Martin}},
  \bibinfo{author}{\bibfnamefont{A.}~\bibnamefont{Zettl}},
  \bibnamefont{et~al.}, \bibinfo{journal}{Nature}
  \textbf{\bibinfo{volume}{520}}, \bibinfo{pages}{650} (\bibinfo{year}{2015}).

\bibitem[{\citenamefont{Sui et~al.}(2015)\citenamefont{Sui, Chen, Ma, Shan,
  Tian, Watanabe, Taniguchi, Jin, Yao, Xiao et~al.}}]{Sui2015}
\bibinfo{author}{\bibfnamefont{M.}~\bibnamefont{Sui}},
  \bibinfo{author}{\bibfnamefont{G.}~\bibnamefont{Chen}},
  \bibinfo{author}{\bibfnamefont{L.}~\bibnamefont{Ma}},
  \bibinfo{author}{\bibfnamefont{W.}~\bibnamefont{Shan}},
  \bibinfo{author}{\bibfnamefont{D.}~\bibnamefont{Tian}},
  \bibinfo{author}{\bibfnamefont{K.}~\bibnamefont{Watanabe}},
  \bibinfo{author}{\bibfnamefont{T.}~\bibnamefont{Taniguchi}},
  \bibinfo{author}{\bibfnamefont{X.}~\bibnamefont{Jin}},
  \bibinfo{author}{\bibfnamefont{W.}~\bibnamefont{Yao}},
  \bibinfo{author}{\bibfnamefont{D.}~\bibnamefont{Xiao}}, \bibnamefont{et~al.},
\bibnamefont{et~al.}, \bibinfo{journal}{Nat Phys}
  \textbf{\bibinfo{volume}{11}}, \bibinfo{pages}{1027} (\bibinfo{year}{2015}).


\bibitem[{\citenamefont{Wu et~al.}(2013)\citenamefont{Wu, Huang, Aivazian,
  Ross, Cobden, and Xu}}]{Wu2013}
\bibinfo{author}{\bibfnamefont{S.}~\bibnamefont{Wu}},
  \bibinfo{author}{\bibfnamefont{C.}~\bibnamefont{Huang}},
  \bibinfo{author}{\bibfnamefont{G.}~\bibnamefont{Aivazian}},
  \bibinfo{author}{\bibfnamefont{J.~S.} \bibnamefont{Ross}},
  \bibinfo{author}{\bibfnamefont{D.~H.} \bibnamefont{Cobden}},
  \bibnamefont{and} \bibinfo{author}{\bibfnamefont{X.}~\bibnamefont{Xu}},
  \bibinfo{journal}{ACS Nano} \textbf{\bibinfo{volume}{7}},
  \bibinfo{pages}{2768} (\bibinfo{year}{2013}).

\bibitem[{\citenamefont{Nomura and MacDonald}(2007)}]{Nomura2007}
\bibinfo{author}{\bibfnamefont{K.}~\bibnamefont{Nomura}} \bibnamefont{and}
  \bibinfo{author}{\bibfnamefont{A.~H.} \bibnamefont{MacDonald}},
  \bibinfo{journal}{Phys. Rev. Lett.} \textbf{\bibinfo{volume}{98}},
  \bibinfo{pages}{076602} (\bibinfo{year}{2007}).

\bibitem[{\citenamefont{Adam et~al.}(2007)\citenamefont{Adam, Hwang, Galitski,
  and Das~Sarma}}]{Adam2007}
\bibinfo{author}{\bibfnamefont{S.}~\bibnamefont{Adam}},
  \bibinfo{author}{\bibfnamefont{E.~H.} \bibnamefont{Hwang}},
  \bibinfo{author}{\bibfnamefont{V.~M.} \bibnamefont{Galitski}},
  \bibnamefont{and}
  \bibinfo{author}{\bibfnamefont{S.}~\bibnamefont{Das~Sarma}},
  \bibinfo{journal}{Proc. Natl. Acad. Sci.} \textbf{\bibinfo{volume}{104}},
  \bibinfo{pages}{18392} (\bibinfo{year}{2007}).

\bibitem[{\citenamefont{Hwang et~al.}(2007)\citenamefont{Hwang, Adam, and
  Das~Sarma}}]{Hwang2007}
\bibinfo{author}{\bibfnamefont{E.~H.} \bibnamefont{Hwang}},
  \bibinfo{author}{\bibfnamefont{S.}~\bibnamefont{Adam}}, \bibnamefont{and}
  \bibinfo{author}{\bibfnamefont{S.}~\bibnamefont{Das~Sarma}},
  \bibinfo{journal}{Phys. Rev. Lett.} \textbf{\bibinfo{volume}{98}},
  \bibinfo{pages}{186806} (\bibinfo{year}{2007}).

\bibitem[{\citenamefont{Chen et~al.}(2009)\citenamefont{Chen, Cullen, Jang,
  Fuhrer, and Williams}}]{Chen2009}
\bibinfo{author}{\bibfnamefont{J.-H.} \bibnamefont{Chen}},
  \bibinfo{author}{\bibfnamefont{W.~G.} \bibnamefont{Cullen}},
  \bibinfo{author}{\bibfnamefont{C.}~\bibnamefont{Jang}},
  \bibinfo{author}{\bibfnamefont{M.~S.} \bibnamefont{Fuhrer}},
  \bibnamefont{and} \bibinfo{author}{\bibfnamefont{E.~D.}
  \bibnamefont{Williams}}, \bibinfo{journal}{Phys. Rev. Lett.}
  \textbf{\bibinfo{volume}{102}}, \bibinfo{pages}{236805}
  (\bibinfo{year}{2009}).

\bibitem[{\citenamefont{Wehling et~al.}(2010)\citenamefont{Wehling, Yuan,
  Lichtenstein, Geim, and Katsnelson}}]{Wehling2010}
\bibinfo{author}{\bibfnamefont{T.~O.} \bibnamefont{Wehling}},
  \bibinfo{author}{\bibfnamefont{S.}~\bibnamefont{Yuan}},
  \bibinfo{author}{\bibfnamefont{A.~I.} \bibnamefont{Lichtenstein}},
  \bibinfo{author}{\bibfnamefont{A.~K.} \bibnamefont{Geim}}, \bibnamefont{and}
  \bibinfo{author}{\bibfnamefont{M.~I.} \bibnamefont{Katsnelson}},
  \bibinfo{journal}{Phys. Rev. Lett.} \textbf{\bibinfo{volume}{105}},
  \bibinfo{pages}{056802} (\bibinfo{year}{2010}).

\bibitem[{\citenamefont{Peres}(2010)}]{Peres2010}
\bibinfo{author}{\bibfnamefont{N.~M.~R.} \bibnamefont{Peres}},
  \bibinfo{journal}{Rev. Mod. Phys.} \textbf{\bibinfo{volume}{82}},
  \bibinfo{pages}{2673} (\bibinfo{year}{2010}).

\bibitem[{\citenamefont{McCann et~al.}(2006)\citenamefont{McCann, Kechedzhi,
  Fal'ko, Suzuura, Ando, and Altshuler}}]{Mccann2006}
\bibinfo{author}{\bibfnamefont{E.}~\bibnamefont{McCann}},
  \bibinfo{author}{\bibfnamefont{K.}~\bibnamefont{Kechedzhi}},
  \bibinfo{author}{\bibfnamefont{V.~I.} \bibnamefont{Fal'ko}},
  \bibinfo{author}{\bibfnamefont{H.}~\bibnamefont{Suzuura}},
  \bibinfo{author}{\bibfnamefont{T.}~\bibnamefont{Ando}}, \bibnamefont{and}
  \bibinfo{author}{\bibfnamefont{B.~L.} \bibnamefont{Altshuler}},
  \bibinfo{journal}{Phys. Rev. Lett.} \textbf{\bibinfo{volume}{97}},
  \bibinfo{pages}{146805} (\bibinfo{year}{2006}).

\bibitem[{\citenamefont{Wu et~al.}(2007)\citenamefont{Wu, Li, Song, Berger, and
  de~Heer}}]{Wu2007}
\bibinfo{author}{\bibfnamefont{X.~S.} \bibnamefont{Wu}},
  \bibinfo{author}{\bibfnamefont{X.}~\bibnamefont{Li}},
  \bibinfo{author}{\bibfnamefont{Z.}~\bibnamefont{Song}},
  \bibinfo{author}{\bibfnamefont{C.}~\bibnamefont{Berger}}, \bibnamefont{and}
  \bibinfo{author}{\bibfnamefont{W.~A.} \bibnamefont{de~Heer}},
  \bibinfo{journal}{Phys. Rev. Lett.} \textbf{\bibinfo{volume}{98}},
  \bibinfo{pages}{136801} (\bibinfo{year}{2007}).

\bibitem[{\citenamefont{Tikhonenko et~al.}(2008)\citenamefont{Tikhonenko,
  Horsell, Gorbachev, and Savchenko}}]{Tikhonenko2008}
\bibinfo{author}{\bibfnamefont{F.~V.} \bibnamefont{Tikhonenko}},
  \bibinfo{author}{\bibfnamefont{D.~W.} \bibnamefont{Horsell}},
  \bibinfo{author}{\bibfnamefont{R.~V.} \bibnamefont{Gorbachev}},
  \bibnamefont{and} \bibinfo{author}{\bibfnamefont{A.~K.}
  \bibnamefont{Savchenko}}, \bibinfo{journal}{Phys. Rev. Lett.}
  \textbf{\bibinfo{volume}{100}}, \bibinfo{pages}{056802}
  (\bibinfo{year}{2008}).

\bibitem[{\citenamefont{Pereira et~al.}(2006)\citenamefont{Pereira, Guinea,
  Lopes~dos Santos, Peres, and Castro~Neto}}]{Pereira2006}
\bibinfo{author}{\bibfnamefont{V.~M.} \bibnamefont{Pereira}},
  \bibinfo{author}{\bibfnamefont{F.}~\bibnamefont{Guinea}},
  \bibinfo{author}{\bibfnamefont{J.~M.~B.} \bibnamefont{Lopes~dos Santos}},
  \bibinfo{author}{\bibfnamefont{N.~M.~R.} \bibnamefont{Peres}},
  \bibnamefont{and} \bibinfo{author}{\bibfnamefont{A.~H.}
  \bibnamefont{Castro~Neto}}, \bibinfo{journal}{Phys. Rev. Lett.}
  \textbf{\bibinfo{volume}{96}}, \bibinfo{pages}{036801}
  (\bibinfo{year}{2006}).

\bibitem[{\citenamefont{Wehling et~al.}(2007)\citenamefont{Wehling, Balatsky,
  Katsnelson, Lichtenstein, Scharnberg, and Wiesendanger}}]{Wehling2007}
\bibinfo{author}{\bibfnamefont{T.~O.} \bibnamefont{Wehling}},
  \bibinfo{author}{\bibfnamefont{A.~V.} \bibnamefont{Balatsky}},
  \bibinfo{author}{\bibfnamefont{M.~I.} \bibnamefont{Katsnelson}},
  \bibinfo{author}{\bibfnamefont{A.~I.} \bibnamefont{Lichtenstein}},
  \bibinfo{author}{\bibfnamefont{K.}~\bibnamefont{Scharnberg}},
  \bibnamefont{and}
  \bibinfo{author}{\bibfnamefont{R.}~\bibnamefont{Wiesendanger}},
  \bibinfo{journal}{Phys. Rev. B} \textbf{\bibinfo{volume}{75}},
  \bibinfo{pages}{125425} (\bibinfo{year}{2007}).

\bibitem[{\citenamefont{Brar et~al.}(2011)\citenamefont{Brar, Decker, Solowan,
  Wang, Maserati, Chan, Lee, Girit, Zettl, Louie et~al.}}]{Brar2011}
\bibinfo{author}{\bibfnamefont{V.~W.} \bibnamefont{Brar}},
  \bibinfo{author}{\bibfnamefont{R.}~\bibnamefont{Decker}},
  \bibinfo{author}{\bibfnamefont{H.-M.} \bibnamefont{Solowan}},
  \bibinfo{author}{\bibfnamefont{Y.}~\bibnamefont{Wang}},
  \bibinfo{author}{\bibfnamefont{L.}~\bibnamefont{Maserati}},
  \bibinfo{author}{\bibfnamefont{K.~T.} \bibnamefont{Chan}},
  \bibinfo{author}{\bibfnamefont{H.}~\bibnamefont{Lee}},
  \bibinfo{author}{\bibfnamefont{C.~O.} \bibnamefont{Girit}},
  \bibinfo{author}{\bibfnamefont{A.}~\bibnamefont{Zettl}},
  \bibinfo{author}{\bibfnamefont{S.~G.} \bibnamefont{Louie}},
  \bibnamefont{et~al.}, \bibinfo{journal}{Nat. Phys.}
  \textbf{\bibinfo{volume}{7}}, \bibinfo{pages}{43} (\bibinfo{year}{2011}).

\bibitem[{\citenamefont{Chen et~al.}(2013)\citenamefont{Chen, Shi, Cai, Xu,
  Sun, Wu, and Yu}}]{Chen2013}
\bibinfo{author}{\bibfnamefont{J.}~\bibnamefont{Chen}},
  \bibinfo{author}{\bibfnamefont{T.}~\bibnamefont{Shi}},
  \bibinfo{author}{\bibfnamefont{T.}~\bibnamefont{Cai}},
  \bibinfo{author}{\bibfnamefont{T.}~\bibnamefont{Xu}},
  \bibinfo{author}{\bibfnamefont{L.}~\bibnamefont{Sun}},
  \bibinfo{author}{\bibfnamefont{X.~S.} \bibnamefont{Wu}}, \bibnamefont{and}
  \bibinfo{author}{\bibfnamefont{D.}~\bibnamefont{Yu}}, \bibinfo{journal}{Appl.
  Phys. Lett.} \textbf{\bibinfo{volume}{102}}, \bibinfo{pages}{103107}
  (\bibinfo{year}{2013}).

\bibitem[{\citenamefont{Nakada and Ishii}(2011)}]{Nakada2011}
\bibinfo{author}{\bibfnamefont{K.}~\bibnamefont{Nakada}} \bibnamefont{and}
  \bibinfo{author}{\bibfnamefont{A.}~\bibnamefont{Ishii}},
  \bibinfo{journal}{Solid State Commun.} \textbf{\bibinfo{volume}{151}},
  \bibinfo{pages}{13} (\bibinfo{year}{2011}).

\bibitem[{\citenamefont{Lucchese et~al.}(2010)\citenamefont{Lucchese, Stavale,
  Ferreira, Vilani, Moutinho, Capaz, Achete, and Jorio}}]{Lucchese2010}
\bibinfo{author}{\bibfnamefont{M.}~\bibnamefont{Lucchese}},
  \bibinfo{author}{\bibfnamefont{F.}~\bibnamefont{Stavale}},
  \bibinfo{author}{\bibfnamefont{E.~M.} \bibnamefont{Ferreira}},
  \bibinfo{author}{\bibfnamefont{C.}~\bibnamefont{Vilani}},
  \bibinfo{author}{\bibfnamefont{M.}~\bibnamefont{Moutinho}},
  \bibinfo{author}{\bibfnamefont{R.~B.} \bibnamefont{Capaz}},
  \bibinfo{author}{\bibfnamefont{C.}~\bibnamefont{Achete}}, \bibnamefont{and}
  \bibinfo{author}{\bibfnamefont{A.}~\bibnamefont{Jorio}},
  \bibinfo{journal}{Carbon} \textbf{\bibinfo{volume}{48}}, \bibinfo{pages}{1592
  } (\bibinfo{year}{2010}).

\bibitem[{\citenamefont{Novikov}(2007)}]{Novikov2007}
\bibinfo{author}{\bibfnamefont{D.~S.} \bibnamefont{Novikov}},
  \bibinfo{journal}{Appl. Phys. Lett.} \textbf{\bibinfo{volume}{91}},
  \bibinfo{eid}{102102} (\bibinfo{year}{2007}).

\bibitem[{\citenamefont{Chen et~al.}(2008)\citenamefont{Chen, Jang, Adam,
  Fuhrer, Williams, and Ishigami}}]{Chen2008}
\bibinfo{author}{\bibfnamefont{J.-H.} \bibnamefont{Chen}},
  \bibinfo{author}{\bibfnamefont{C.}~\bibnamefont{Jang}},
  \bibinfo{author}{\bibfnamefont{S.}~\bibnamefont{Adam}},
  \bibinfo{author}{\bibfnamefont{M.}~\bibnamefont{Fuhrer}},
  \bibinfo{author}{\bibfnamefont{E.}~\bibnamefont{Williams}}, \bibnamefont{and}
  \bibinfo{author}{\bibfnamefont{M.}~\bibnamefont{Ishigami}},
  \bibinfo{journal}{Nat. Phys.} \textbf{\bibinfo{volume}{4}},
  \bibinfo{pages}{377} (\bibinfo{year}{2008}).

\bibitem[{\citenamefont{Nanda et~al.}(2012)\citenamefont{Nanda, Sherafati,
  Popovic, and Satpathy}}]{Nanda2012}
\bibinfo{author}{\bibfnamefont{B.~R.~K.} \bibnamefont{Nanda}},
  \bibinfo{author}{\bibfnamefont{M.}~\bibnamefont{Sherafati}},
  \bibinfo{author}{\bibfnamefont{Z.~S.} \bibnamefont{Popovic}},
  \bibnamefont{and} \bibinfo{author}{\bibfnamefont{S.}~\bibnamefont{Satpathy}},
  \bibinfo{journal}{New J. Phys.} \textbf{\bibinfo{volume}{14}},
  \bibinfo{pages}{083004} (\bibinfo{year}{2012}).

\bibitem[{Not()}]{Note1}
\bibinfo{note}{Although defects also introduce $sp^2\sigma$ derived states
  according to \rref{Nanda2012}, these states are far away from the Dirac point
  and half-filled (with three electrons). Thus, they do not contribute to the
  net charge of a vacancy.}

\bibitem[{\citenamefont{Ugeda et~al.}(2010)\citenamefont{Ugeda, Brihuega,
  Guinea, and G\'omez-Rodr\'iguez}}]{Ugeda2010}
\bibinfo{author}{\bibfnamefont{M.~M.} \bibnamefont{Ugeda}},
  \bibinfo{author}{\bibfnamefont{I.}~\bibnamefont{Brihuega}},
  \bibinfo{author}{\bibfnamefont{F.}~\bibnamefont{Guinea}}, \bibnamefont{and}
  \bibinfo{author}{\bibfnamefont{J.~M.} \bibnamefont{G\'omez-Rodr\'iguez}},
  \bibinfo{journal}{Phys. Rev. Lett.} \textbf{\bibinfo{volume}{104}},
  \bibinfo{pages}{096804} (\bibinfo{year}{2010}).

\bibitem[{\citenamefont{Liu et~al.}(2014)\citenamefont{Liu, Weinert, and
  Li}}]{Liu2014a}
\bibinfo{author}{\bibfnamefont{Y.}~\bibnamefont{Liu}},
  \bibinfo{author}{\bibfnamefont{M.}~\bibnamefont{Weinert}}, \bibnamefont{and}
  \bibinfo{author}{\bibfnamefont{L.}~\bibnamefont{Li}},
  \bibinfo{journal}{Nanotechnology} \textbf{\bibinfo{volume}{26}},
  \bibinfo{pages}{035702} (\bibinfo{year}{2014}).

\bibitem[{\citenamefont{Chan et~al.}(2008)\citenamefont{Chan, Neaton, and
  Cohen}}]{Chan2008}
\bibinfo{author}{\bibfnamefont{K.~T.} \bibnamefont{Chan}},
  \bibinfo{author}{\bibfnamefont{J.~B.} \bibnamefont{Neaton}},
  \bibnamefont{and} \bibinfo{author}{\bibfnamefont{M.~L.} \bibnamefont{Cohen}},
  \bibinfo{journal}{Phys. Rev. B} \textbf{\bibinfo{volume}{77}},
  \bibinfo{pages}{235430} (\bibinfo{year}{2008}).

\bibitem[{\citenamefont{{Castro Neto} and Guinea}(2009)}]{CastroNeto2009}
\bibinfo{author}{\bibfnamefont{A.~H.} \bibnamefont{{Castro Neto}}}
  \bibnamefont{and} \bibinfo{author}{\bibfnamefont{F.}~\bibnamefont{Guinea}},
  \bibinfo{journal}{Phys. Rev. Lett.} \textbf{\bibinfo{volume}{103}},
  \bibinfo{pages}{026804} (\bibinfo{year}{2009}).

\bibitem[{\citenamefont{Weeks et~al.}(2011)\citenamefont{Weeks, Hu, Alicea,
  Franz, and Wu}}]{Weeks2011}
\bibinfo{author}{\bibfnamefont{C.}~\bibnamefont{Weeks}},
  \bibinfo{author}{\bibfnamefont{J.}~\bibnamefont{Hu}},
  \bibinfo{author}{\bibfnamefont{J.}~\bibnamefont{Alicea}},
  \bibinfo{author}{\bibfnamefont{M.}~\bibnamefont{Franz}}, \bibnamefont{and}
  \bibinfo{author}{\bibfnamefont{R.}~\bibnamefont{Wu}}, \bibinfo{journal}{Phys.
  Rev. X} \textbf{\bibinfo{volume}{1}}, \bibinfo{pages}{021001}
  (\bibinfo{year}{2011}).

\bibitem[{\citenamefont{Jiang et~al.}(2012)\citenamefont{Jiang, Qiao, Liu, Shi,
  and Niu}}]{Jiang2012}
\bibinfo{author}{\bibfnamefont{H.}~\bibnamefont{Jiang}},
  \bibinfo{author}{\bibfnamefont{Z.}~\bibnamefont{Qiao}},
  \bibinfo{author}{\bibfnamefont{H.}~\bibnamefont{Liu}},
  \bibinfo{author}{\bibfnamefont{J.}~\bibnamefont{Shi}}, \bibnamefont{and}
  \bibinfo{author}{\bibfnamefont{Q.}~\bibnamefont{Niu}},
  \bibinfo{journal}{Phys. Rev. Lett.} \textbf{\bibinfo{volume}{109}},
  \bibinfo{pages}{116803} (\bibinfo{year}{2012}).

\bibitem[{\citenamefont{Ferreira et~al.}(2014)\citenamefont{Ferreira,
  Rappoport, Cazalilla, and Castro~Neto}}]{Ferreira2014}
\bibinfo{author}{\bibfnamefont{A.}~\bibnamefont{Ferreira}},
  \bibinfo{author}{\bibfnamefont{T.~G.} \bibnamefont{Rappoport}},
  \bibinfo{author}{\bibfnamefont{M.~A.} \bibnamefont{Cazalilla}},
  \bibnamefont{and} \bibinfo{author}{\bibfnamefont{A.~H.}
  \bibnamefont{Castro~Neto}}, \bibinfo{journal}{Phys. Rev. Lett.}
  \textbf{\bibinfo{volume}{112}}, \bibinfo{pages}{066601}
  (\bibinfo{year}{2014}).

\bibitem[{\citenamefont{Cresti et~al.}(2014)\citenamefont{Cresti, {Van Tuan},
  Soriano, Cummings, and Roche}}]{Cresti2014}
\bibinfo{author}{\bibfnamefont{A.}~\bibnamefont{Cresti}},
  \bibinfo{author}{\bibfnamefont{D.}~\bibnamefont{{Van Tuan}}},
  \bibinfo{author}{\bibfnamefont{D.}~\bibnamefont{Soriano}},
  \bibinfo{author}{\bibfnamefont{A.~W.} \bibnamefont{Cummings}},
  \bibnamefont{and} \bibinfo{author}{\bibfnamefont{S.}~\bibnamefont{Roche}},
  \bibinfo{journal}{Phys. Rev. Lett.} \textbf{\bibinfo{volume}{113}},
  \bibinfo{pages}{246603} (\bibinfo{year}{2014}).

\bibitem[{\citenamefont{Sevincli et~al.}(2008)\citenamefont{Sevincli, Topsakal,
  Durgun, and Ciraci}}]{Sevincli2008}
\bibinfo{author}{\bibfnamefont{H.}~\bibnamefont{Sevincli}},
  \bibinfo{author}{\bibfnamefont{M.}~\bibnamefont{Topsakal}},
  \bibinfo{author}{\bibfnamefont{E.}~\bibnamefont{Durgun}}, \bibnamefont{and}
  \bibinfo{author}{\bibfnamefont{S.}~\bibnamefont{Ciraci}},
  \bibinfo{journal}{Phys. Rev. B} \textbf{\bibinfo{volume}{77}},
  \bibinfo{pages}{195434} (\bibinfo{year}{2008}).

\bibitem[{\citenamefont{Qiao et~al.}(2010)\citenamefont{Qiao, Yang, Feng, Tse,
  Ding, Yao, Wang, and Niu}}]{Qiao2010}
\bibinfo{author}{\bibfnamefont{Z.~H.} \bibnamefont{Qiao}},
  \bibinfo{author}{\bibfnamefont{S.~Y.~A.} \bibnamefont{Yang}},
  \bibinfo{author}{\bibfnamefont{W.~X.} \bibnamefont{Feng}},
  \bibinfo{author}{\bibfnamefont{W.~K.} \bibnamefont{Tse}},
  \bibinfo{author}{\bibfnamefont{J.}~\bibnamefont{Ding}},
  \bibinfo{author}{\bibfnamefont{Y.~G.} \bibnamefont{Yao}},
  \bibinfo{author}{\bibfnamefont{J.}~\bibnamefont{Wang}}, \bibnamefont{and}
  \bibinfo{author}{\bibfnamefont{Q.}~\bibnamefont{Niu}},
  \bibinfo{journal}{Phys. Rev. B} \textbf{\bibinfo{volume}{82}},
  \bibinfo{pages}{161414} (\bibinfo{year}{2010}).

\bibitem[{\citenamefont{Cao et~al.}(2010)\citenamefont{Cao, Wu, Jiang, and
  Cheng}}]{Cao2010}
\bibinfo{author}{\bibfnamefont{C.}~\bibnamefont{Cao}},
  \bibinfo{author}{\bibfnamefont{M.}~\bibnamefont{Wu}},
  \bibinfo{author}{\bibfnamefont{J.}~\bibnamefont{Jiang}}, \bibnamefont{and}
  \bibinfo{author}{\bibfnamefont{H.-P.} \bibnamefont{Cheng}},
  \bibinfo{journal}{Phys. Rev. B} \textbf{\bibinfo{volume}{81}},
  \bibinfo{pages}{205424} (\bibinfo{year}{2010}).

\bibitem[{\citenamefont{Virgus et~al.}(2014)\citenamefont{Virgus, Purwanto,
  Krakauer, and Zhang}}]{Virgus2014}
\bibinfo{author}{\bibfnamefont{Y.}~\bibnamefont{Virgus}},
  \bibinfo{author}{\bibfnamefont{W.}~\bibnamefont{Purwanto}},
  \bibinfo{author}{\bibfnamefont{H.}~\bibnamefont{Krakauer}}, \bibnamefont{and}
  \bibinfo{author}{\bibfnamefont{S.}~\bibnamefont{Zhang}},
  \bibinfo{journal}{Phys. Rev. Lett.} \textbf{\bibinfo{volume}{113}},
  \bibinfo{pages}{175502} (\bibinfo{year}{2014}).

\bibitem[{\citenamefont{Pi et~al.}(2010)\citenamefont{Pi, Han, McCreary,
  Swartz, Li, and Kawakami}}]{Pi2010}
\bibinfo{author}{\bibfnamefont{K.}~\bibnamefont{Pi}},
  \bibinfo{author}{\bibfnamefont{W.}~\bibnamefont{Han}},
  \bibinfo{author}{\bibfnamefont{K.~M.} \bibnamefont{McCreary}},
  \bibinfo{author}{\bibfnamefont{A.~G.} \bibnamefont{Swartz}},
  \bibinfo{author}{\bibfnamefont{Y.}~\bibnamefont{Li}}, \bibnamefont{and}
  \bibinfo{author}{\bibfnamefont{R.~K.} \bibnamefont{Kawakami}},
  \bibinfo{journal}{Phys. Rev. Lett.} \textbf{\bibinfo{volume}{104}},
  \bibinfo{pages}{187201} (\bibinfo{year}{2010}).

\bibitem[{\citenamefont{Swartz et~al.}(2013)\citenamefont{Swartz, Chen,
  McCreary, Odenthal, Han, and Kawakami}}]{Swartz2013}
\bibinfo{author}{\bibfnamefont{A.~G.} \bibnamefont{Swartz}},
  \bibinfo{author}{\bibfnamefont{J.-R.} \bibnamefont{Chen}},
  \bibinfo{author}{\bibfnamefont{K.~M.} \bibnamefont{McCreary}},
  \bibinfo{author}{\bibfnamefont{P.~M.} \bibnamefont{Odenthal}},
  \bibinfo{author}{\bibfnamefont{W.}~\bibnamefont{Han}}, \bibnamefont{and}
  \bibinfo{author}{\bibfnamefont{R.~K.} \bibnamefont{Kawakami}},
  \bibinfo{journal}{Phys. Rev. B} \textbf{\bibinfo{volume}{87}},
  \bibinfo{pages}{075455} (\bibinfo{year}{2013}).

\bibitem[{\citenamefont{Jia et~al.}(2015)\citenamefont{Jia, Yan, Niu, Han, Zhu,
  Yu, and Wu}}]{Jia2015}
\bibinfo{author}{\bibfnamefont{Z.}~\bibnamefont{Jia}},
  \bibinfo{author}{\bibfnamefont{B.}~\bibnamefont{Yan}},
  \bibinfo{author}{\bibfnamefont{J.}~\bibnamefont{Niu}},
  \bibinfo{author}{\bibfnamefont{Q.}~\bibnamefont{Han}},
  \bibinfo{author}{\bibfnamefont{R.}~\bibnamefont{Zhu}},
  \bibinfo{author}{\bibfnamefont{D.}~\bibnamefont{Yu}}, \bibnamefont{and}
  \bibinfo{author}{\bibfnamefont{X.}~\bibnamefont{Wu}}, \bibinfo{journal}{Phys.
  Rev. B} \textbf{\bibinfo{volume}{91}}, \bibinfo{pages}{085411}
  (\bibinfo{year}{2015}).

\bibitem[{\citenamefont{Chandni et~al.}(2015)\citenamefont{Chandni, Henriksen,
  and Eisenstein}}]{Chandni2015}
\bibinfo{author}{\bibfnamefont{U.}~\bibnamefont{Chandni}},
  \bibinfo{author}{\bibfnamefont{E.~A.} \bibnamefont{Henriksen}},
  \bibnamefont{and} \bibinfo{author}{\bibfnamefont{J.~P.}
  \bibnamefont{Eisenstein}}, \bibinfo{journal}{Phys. Rev. B}
  \textbf{\bibinfo{volume}{91}}, \bibinfo{pages}{245402}
  (\bibinfo{year}{2015}).

\end{thebibliography}
\end{document}